\documentclass[aps,prl,preprint,superscriptaddress]{revtex4-1}
\usepackage{natbib}
\usepackage{amsmath}
\usepackage[toc,page]{appendix}
\usepackage{bm}
\usepackage{ amssymb } 
\usepackage{graphicx}
\usepackage {epstopdf}
\usepackage{csquotes}

\begin{document}


\title{Quantum Lifetime in
Ultra-High Quality GaAs Quantum Wells: Relationship to $\Delta_{5/2}$ and Impact of Density Fluctuations
}


\author{Q. Qian}
\affiliation{Department of Physics and Astronomy, Purdue University}
\affiliation{Station Q Purdue, Purdue University}

\author{J. Nakamura}
\affiliation{Department of Physics and Astronomy, Purdue University}
\affiliation{Station Q Purdue, Purdue University}

\author{S. Fallahi}
\affiliation{Department of Physics and Astronomy, Purdue University}
\affiliation{Station Q Purdue, Purdue University}
\affiliation{Birck Nanotechnology Center, Purdue University}

\author{G. C. Gardner}
\affiliation{Station Q Purdue, Purdue University}
\affiliation{Birck Nanotechnology Center, Purdue University}
\affiliation{School of Materials Engineering, Purdue University}

\author{J. D. Watson}
\affiliation{Department of Physics and Astronomy, Purdue University}
\affiliation{Birck Nanotechnology Center, Purdue University}

\author{S. L\"uscher}
\affiliation{ Department of Physics and Astronomy, University of British Columbia, Vancouver, BC V6T 1Z1, Canada }
\affiliation{Quantum Matter Institute, University of British Columbia, Vancouver, BC V6T 1Z4, Canada}

\author{J. A. Folk}
\affiliation{ Department of Physics and Astronomy, University of British Columbia, Vancouver, BC V6T 1Z1, Canada }
\affiliation{Quantum Matter Institute, University of British Columbia, Vancouver, BC V6T 1Z4, Canada}

\author{G. A. Cs\'athy}
\affiliation{Department of Physics and Astronomy, Purdue University}

\author{M. J. Manfra}
\email[]{mmanfra@purdue.edu}
\affiliation{Department of Physics and Astronomy, Purdue University}
\affiliation{Station Q Purdue, Purdue University}
\affiliation{Birck Nanotechnology Center, Purdue University}

\affiliation{School of Materials Engineering, Purdue University}
\affiliation{School of Electrical and Computer Engineering, Purdue University}

\date{\today}
\begin{abstract}
We consider quantum lifetime derived from low-field Shubnikov-de Haas oscillations as a metric of quality of the two-dimensional electron gas in GaAs quantum wells that expresses large excitation gaps in the fractional quantum Hall states of the N=1 Landau level. Analysis indicates two salient features: 1) small density inhomogeneities dramatically impact the amplitude of Shubnikov-de Haas oscillations such that the canonical method (cf. Coleridge, Phys. Rev. B \textbf{44}, 3793) for determination of quantum lifetime substantially
underestimates $\tau_q$ unless density inhomogeneity is explicitly considered; 2) $\tau_q$ does not correlate well with quality as measured by $\Delta_{5/2}$, the excitation gap of the fractional quantum Hall state at 5/2 filling.
\end{abstract}

\maketitle
Improvements in heterostructure design and molecular beam epitaxy (MBE) techniques have made it possible to grow AlGaAs/GaAs heterostructures with low-temperature mobility $\mu$ as high as $35 \times 10^{6}  cm^2/Vs$  
\cite{Gardner2016, Manfrareview, Umansky2009}. Ultra-high quality two-dimensional electron gases (2DEGs) provide a platform to study the most fragile fractional quantum Hall states (FQHSs) in the N=1 Landau level (LL), including the putative non-Abelian $\nu=5/2$ and $\nu=12/5$ FQHS.  The existence of $\nu=5/2$ and $\nu=12/5$ states present fundamental challenges to our understanding of correlations in the fractional quantum Hall regime and may provide a platform for exploration of exotic braiding statistics. However, it is often difficult to assess the quality of a given sample by measurement of mobility alone \cite{Neubler2010, Gamez, Manfrareview, DasSarma2014, Pan2011, Umansky2009}. It has been proposed that the quantum scattering time (or quantum lifetime), $\tau_q$, may be a better predictor of the strength of FQHSs at low temperatures and can be used to quantify disorder-induced Landau level broadening \cite{DasSarma2014}. In this study, we investigate the relationship between  $\tau_q$ and the strength of $\nu=5/2$ FQHS in the ultra-high quality GaAs quantum wells. 

Mobility can be recast as a lifetime, $\tau_t=m^*\mu/e$, that depends on the electron effective mass, $m^*$, the mobility, $\mu$, and the electronic charge, $e$.  $\tau_t$ is particularly sensitive to large-angle scattering.  This can be seen in its defining integral: 
\begin{equation}
\frac{1}{\tau_t}=\frac{m^*}{\pi\hbar^3}\int_0^\pi \mathrm|V_q|^2(1-\cos\theta)\,\mathrm{d}\theta
\end{equation}
with $|V_q|$ being the probability of scattering through an angle $\theta$  from a state $k$  to a state $k^{'}$ on the Fermi surface.   Note $q=2k_F\sin({\theta\over2})$ and the Fermi wave-vector $k_F=\sqrt{2\pi n}$. 
The factor of $(1-\cos\theta)$ in the integrand results in reduced weighting of small-angle scattering.  Historically, mobility has been the primary metric of 2DEG quality.

The quantum lifetime is another measure of 2DEG quality that is often used in conjunction with mobility measurements to determine dominant scattering mechanisms \cite{DasSarma1985, Hamilton2013,Manfra2004, MacLeod2009}.
Unlike $\tau_t$, the quantum lifetime weighs all scattering events equally.  The quantum lifetime is defined as \cite{DasSarma1985}:
\begin{equation}
\frac{1}{\tau_q} = \frac{m^*}{\pi\hbar^3}\int_0^\pi \mathrm|V_q|^2\,\mathrm{d}\theta
\end{equation}
 It measures the mean time a carrier remains in a particular momentum eigenstate before being scattered into a different state.  Extraction of $\tau_q$ is usually accomplished with transport measurements through analysis of low magnetic field Shubnikov-de Haas (SdH) oscillations. 

The density of states $g(\epsilon)$ of a 2DEG becomes oscillatory at low magnetic field \cite{Lifshitz1955,Ando1974,AndoReview}.  The functional form of $\Delta g(\epsilon)/g_0$ was derived by Isihara and Smrcka \cite{Isihara}
\begin{equation}
\frac{\Delta g}{g_0}=2\sum_{s=1}^{\infty}\exp(-\frac{\pi s}{\omega_c \tau_q})\cos(\frac{2\pi \epsilon s}{\hbar \omega_c}-s\pi)
\label{eq3}
\end{equation}
where $\omega_c=eB/m^*$ is the cyclotron frequency, $\epsilon$ is the electron energy, and $g_0$ is the 2D density of states at zero magnetic field. Here the quantum lifetime is related to the width of disorder-broadened Landau levels ($\Gamma$) through the relationship $\tau_q=\hbar/2\Gamma$.  
At small magnetic fields, $\omega_c\tau_q\sim1$; retaining only the $s=1$ term in the density of states, the functional form for SdH oscillations can be written as:
\begin{equation}
\Delta R_{xx}=4R_o exp(\frac{-\pi}{\omega_{c}\tau_{q}})cos(\frac{2\hbar\pi^2  n}{m^*\omega_{c}}-\pi)\chi(T)
\label{eq1}
\end{equation}
where $R_{o}$ is the zero field resistance, $n$ is the 2DEG density and $\chi(T)$, a thermal damping factor, is given by
$\chi(T)=(2\pi^{2}kT/\hbar\omega_{c})/sinh(2\pi^{2}kT/\hbar\omega_{c})$.

In a formalism codified by Coleridge {\it et al.} \cite{Coleridge1989, Coleridge1991} $\tau_q$ can be related to the amplitude of developing Shubnikov-de Haas oscillations by the expression:
\begin{equation}
\Delta R=4R_{o}\chi(T)exp(-\pi/\omega_{c}\tau_{q})
\label{eq4}
\end{equation}
Thus $1/\tau_{q}$ can be extracted directly from the slope of $\Delta R/4R_{o}/\chi(T)$ plotted versus $1/B$ in natural logarithm scale, also known as a Dingle plot.  Assuming a homogeneous 2DEG, data plotted in this manner should fall on a straight line with a $1/B=0$ intercept of 4. As discussed below, this assumption is often far from valid for the highest quality 2DEGs available today, requiring a more sophisticated application of the Dingle plot formalism in order to extract $1/\tau_{q}$.

We present measurements on $\textit{in situ}$ back-gated 2DEGs grown by MBE. The 2DEG
resides in 30nm GaAs quantum well bounded by Al$_{0.24}$Ga$_{0.76}$As barriers. Charge transfer to the quantum well is accomplished by $\delta$-doping silicon in narrow GaAs layers flanked by
pure AlAs layers placed 66nm above the principal 30nm GaAs quantum well. This design has been shown to yield the largest gap energy for the $\nu=5/2$ FQHS \cite{Waston, Manfrareview, Deng1, Deng2, Josh}. The $\textit{in situ}$ gate is an $n^{+}$ GaAs layer situated 850 nm below the bottom interface of the quantum well. Leakage from gate to 2DEG is minimized by a 830 nm GaAs/AlAs superlattice in the intervening layer . The 2DEG density can typically be tuned from depletion to $4 \times 10^{11}/ cm^2$ without significant gate leakage (for larger gate voltages, leakage current exceeds 10pA which causes excessive electron heating). We have measured three devices on the same chip, sharing a global back gate. Each device is a 1mm by 1mm lithographically-defined Van der Pauw square with eight contacts on the edges. Most of the data was taken after briefly illuminating the samples with a red LED, although one exception to this is noted in the text. This particular wafer was chosen because it exhibits the largest $\Delta_{5/2}$=0.625K reported to date.
Details of other properties of the devices can be found in Ref. \cite{Waston}.

\begin{figure}[t]
\def\ffile{Fig1.pdf}
\centering
\includegraphics[width=0.6\textwidth]{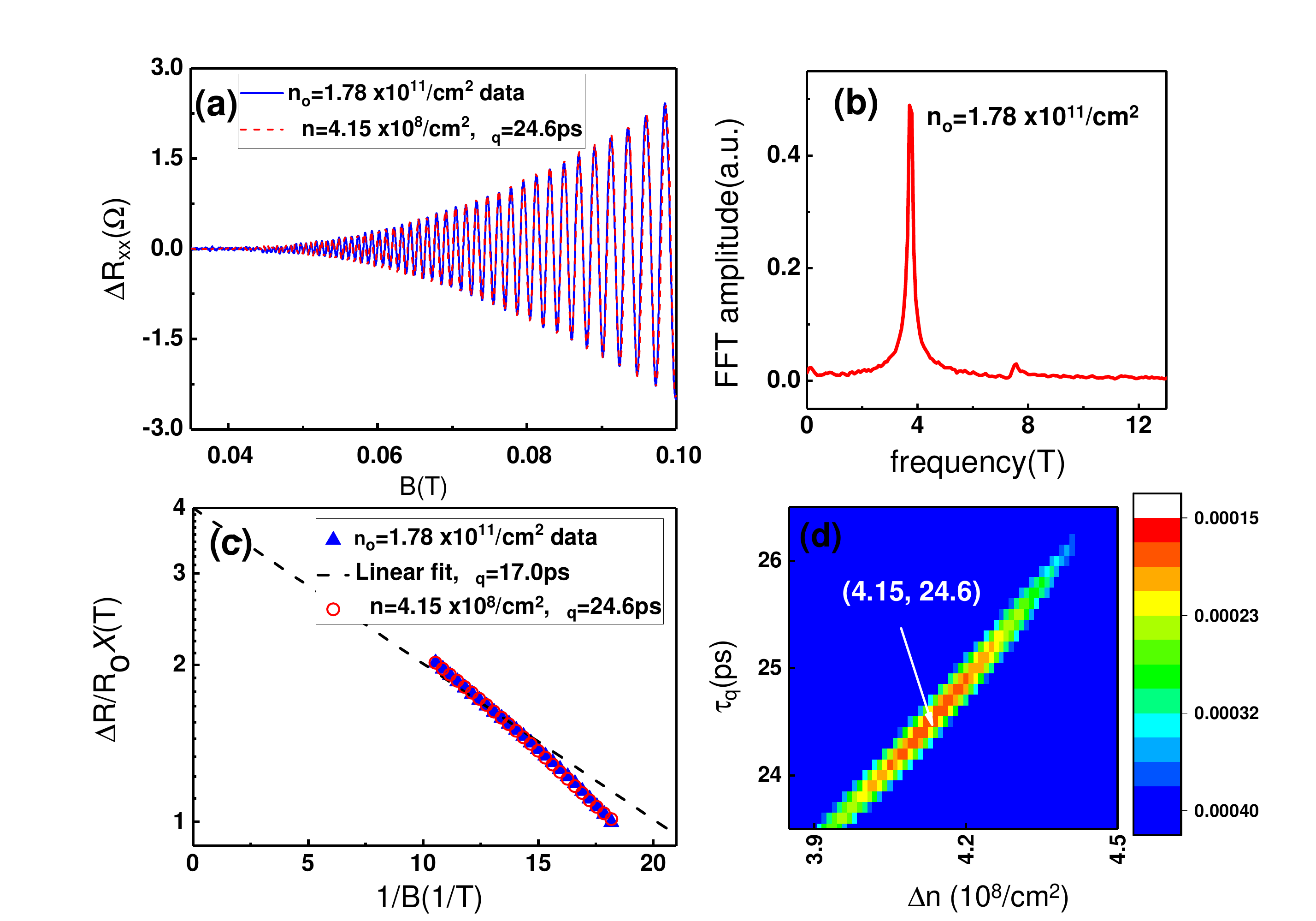}
\caption{\label{methodexample} (color online). Impact of density inhomogeneity on low field transport measured at T=0.3K. (a) Magnetoresistance as a function of B  after background subtraction for nominal 2DEG density $n_{o}=1.78\times10^{11}/cm^{2}$ and simulated trace with $\Delta n/n_o\sim0.25\%$ density inhomogeneity and $\tau_{q}$=24.6ps. 
(b) Density spectrum obtained through a FFT of $\Delta R_{xx}$ vs 1/B.  (c) Dingle plots from data and simulated trace shown in (a). (d) Two-dimensional plot of the fit quality for various combinations of $\Delta n$ and $\tau_q$ for data shown in (a). The error is minimized at $\Delta n=4.15\times 10^{8}$cm$^{-2}$ and $\tau_q = 24.6$ps.}
\end{figure}
Fig. \ref{methodexample}a shows the magnetoresistance of a device at zero gate bias after subtraction of a smooth background. The resistance is measured by monitoring the voltage drop along one edge of the sample while driving current between two contacts at the center of opposing faces of the square.  The amplitude of the oscillations appears to be described by a single envelope function and no beating is observed.  The density spectrum obtained from a fast Fourier transform (FFT) of $\Delta R_{xx}$ vs. 1/B is shown in Fig. \ref{methodexample}b.  Only a narrow fundamental peak associated with the nominal 2DEG density and exact higher-order harmonics are observed \cite{Endo}, indicating the sample does not suffer from gross density inhomogeneity or from two or more regions with distinct densities \cite{Grayson2015}.  As we demonstrate below, however, it is likely that small density inhomogeneities in these samples limit the onset of SdH oscillations at low magnetic field \cite{Sheyum}.

Fig. \ref{methodexample}c is a Dingle plot for the data in Fig. \ref{methodexample}a. A single-parameter least square fit of the data between 55mT and 95mT yields a quantum lifetime $\tau_q$=17ps.  However, the data points clearly deviate from the straight line expected for a sample with homogeneous density \cite{Coleridge1991}.

It is known that even slight density inhomogeneities or gradients can impact transport at high magnetic fields in the quantum Hall regime \cite{Horst1,steve1,Horst2,Ilan2006}. As we show below, minute levels of inhomogeneity can also dominate the low field magnetoresistance when small angle scattering has been strongly suppressed by strong screening of remote scattering centers. For samples shown in Fig. \ref{methodexample}a, the onset of SdH oscillations is around 45mT at T=0.3K, corresponding to filling factor $\nu\sim$ 165, where $\nu={nhc}/{eB}$. At a qualitative level, this onset field could correspond to Landau level broadening associated with $\tau_q$.  On the other hand, density inhomogeneity on the order of 1/$\nu$ $ \sim$ 1/165 $\sim$ 0.5$\%$ will preclude observation of well-defined oscillations at lower magnetic field even in the limit of infinite $\tau_q$.

In order to model the effect of inhomogeneities quantitatively, we assume a Gaussian distribution of densities $n_i$ around nominal density $n_o$ with standard deviation $\Delta n$. 
The 2DEG density distribution is then described by
\begin{equation}
g(n_i)=\frac{1}{\Delta n \sqrt{2\pi}}e^{-\frac{1}{2}(\frac{n_i-n_o}{\Delta n})^2}
\label{A1}
\end{equation}
where $n_o$ is the nominal 2DEG density obtained from FFT spectrum of $\Delta R_{xx}$ vs. 1/B.
For computational purposes the densities are discretized and evenly spaced, and the weight given to each discrete density $n_i$ is denoted as $P(n_i)$ \cite{P}. It is assumed that each density carries the same quantum lifetime $\tau_{q}$.
The resultant magnetotransport at low field then can be expressed as the sum of the distribution of all partial SdH oscillations
\begin{equation}
\Delta R_{xx}=4R_o \sum\limits_{i=1}^m P(n_i) exp(\frac{-\pi}{\omega_{c}\tau_{q}})cos(\frac{2\hbar\pi^2  n_i}{m^*\omega_{c}}-\pi)\chi(T)
\label{A4}
\end{equation}
 This sum of a spread of oscillation frequencies (expressed in 1/B) damps the net oscillation amplitude heavily at small B and results in curvature in a Dingle plot.  For samples with low scattering rates (that is, very high quality and long $\tau_q$), the effect can be enormous. Throughout this paper, Dingle plots are superimposed with simulation results after a standard least squares regression analysis to obtain the best-fit $\Delta n$ and $\tau_q$.  In Fig. \ref{methodexample}d we plot the error incurred for various combinations of $\Delta n$ and $\tau_q$, for the data in  Fig. \ref{methodexample}a.  The error is minimized at $\Delta n=4.15\times 10^{8}$/cm$^{2}$ and $\tau_q = 24.6$ps. 
Next we compare our data to the simulated  SdH oscillations (Fig.\ref{methodexample} a) and Dingle plot (Fig.\ref{methodexample} c) with a quantum lifetime $\tau_q$=24.6ps and density fluctuations $\Delta n/n_o\sim0.25\%$. The excellent overlap between data and simulation reveals that the inclusion of this physically reasonable amount of density variation is necessary to reproduce the actual data. The 24.6ps quantum lifetime obtained after properly accounting for density inhomogeneity is $45\%$ higher than the value of 17ps obtained from a naive linear fit.   

\begin{figure}[t]
\def\ffile{Fig2.pdf}
\centering
\includegraphics[width=0.5\textwidth]{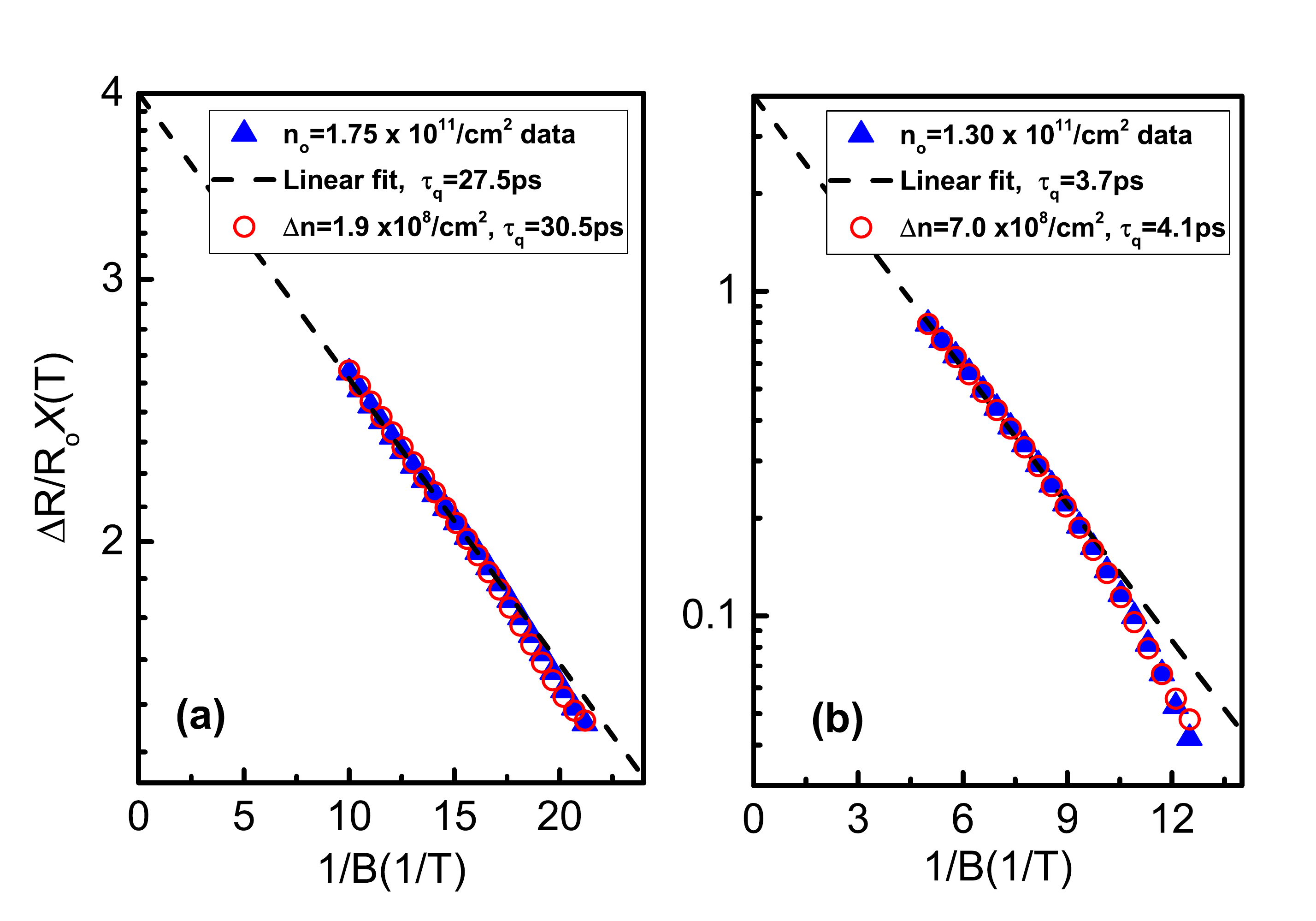}
\caption{\label{300mK} (color online). Dingle plots from data taken at T=0.3K and simulations. (a) Back gate 2DEG with nominal 2DEG density $n_{o}=1.75\times10^{11}/cm^{2}$. (b) A 91nm deep, single heterojunction 2DEG, $n_{o}=1.30\times10^{11}/cm^{2}$.}
\end{figure}

The inaccuracy of $\tau_q$ extracted from a linear fit to the Dingle plot is exacerbated at larger $\tau_q$ and lower temperature. Data in Fig. \ref{300mK}a is from the back-gated device (the same chip as discussed in Fig. \ref{methodexample}), while data shown in Fig. \ref{300mK}b is from a 2DEG utilizing a different heterostructure known to have shorter $\tau_q$.  This lower quality 2DEG is formed at an Al$_{0.36}$Ga$_{0.64}$As/GaAs single heterojunction located 91nm below the surface with silicon uniformly doped in the Al$_{0.36}$Ga$_{0.64}$As layer. The low field data is collected at T=0.3K.
For the single heterojunction wafer in Fig. \ref{300mK}b we find density inhomogeneity  $\Delta n/n_o\sim0.5\%$ and we find that after accounting for the effect of density inhomogeneity, the calculated  $\tau_{q}$ increases by $10\%$. For the back-gated wafer in Fig. \ref{300mK}a we find density inhomogeneity $\Delta n/n_o\sim0.1\%$ (5 times smaller than for the single heterojunction wafer) and we find that the calculated $\tau_{q}$  also increases by $10\%$ when accounting for density inhomogeneity (the same fractional change as for the single heterojunction wafer). Thus, we conclude that accounting for density inhomogeneity has a more pronounced effect on the correction to $\tau_{q}$ for the higher quality back-gated sample than for the single heterojunction sample, since a much smaller density inhomogeneity in the back-gated sample results in the same percentage correction to $\tau_{q}$. 
 \begin{figure}[t]
\def\ffile{Fig3.pdf}
\centering
\includegraphics[width=0.5\textwidth]{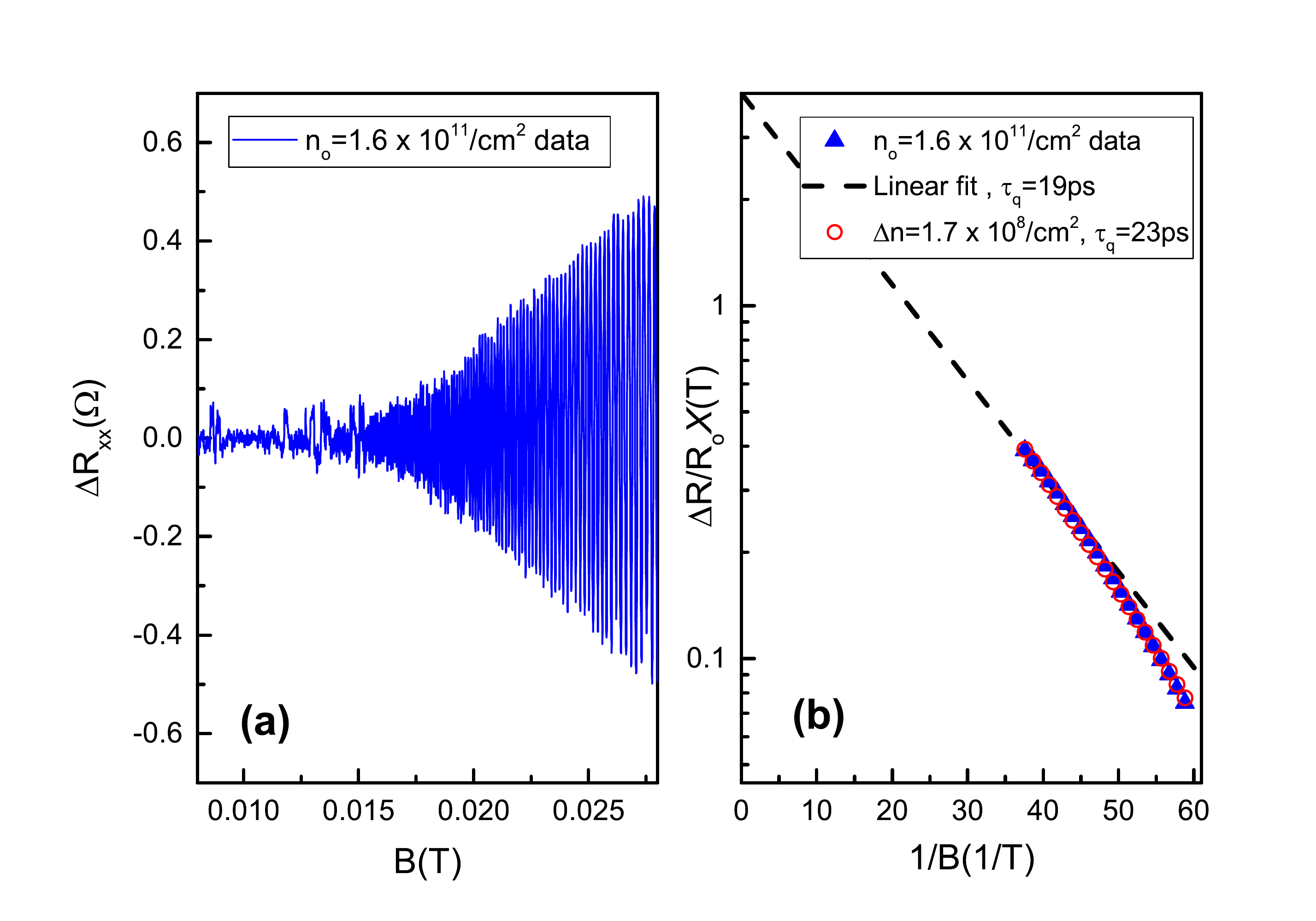}
\caption{\label{10mK}(color online). Low field transport data of back-gated 2DEG with nominal 2DEG density $n_{o}=1.60\times10^{11}/cm^{2}$ taken at T=10mK. (a) Magnetoresistance as a function of B after background subtraction. (b) Dingle plots of experimental data and simulation.}
\end{figure}
The impact of temperature is explored in Fig. \ref{10mK}. Here the back-gated sample is cooled in a dilution refrigerator to T=10mK. The thermal damping effect is largely suppressed at this temperature and as a result the SdH oscillation onset moves to lower field ($\sim$15mT), as shown in Fig. \ref{10mK}a. Now we compare Fig. \ref{300mK}a and  Fig. \ref{10mK}b; data for each plot is from samples with the same heterostructure but measured at different temperatures. They both have $\Delta n/n_o\sim0.1\%$ density inhomogeneity, but the correction to $\tau_{q}$ is $\sim20\%$ for the sample measured at T=10mK  compared to $10\%$ at T=0.3K; also, the curvature of the Dingle plot is more visible in the lower temperature data. Both larger $\tau_{q}$ and lower measurement temperature move the onset of SdH oscillation to lower field, or equivalently higher filling factor $\nu$, where density inhomogeneity will have more impact. 

\begin{figure}[t]
\def\ffile{Fig4.pdf}
\centering
\includegraphics[width=0.5\textwidth]{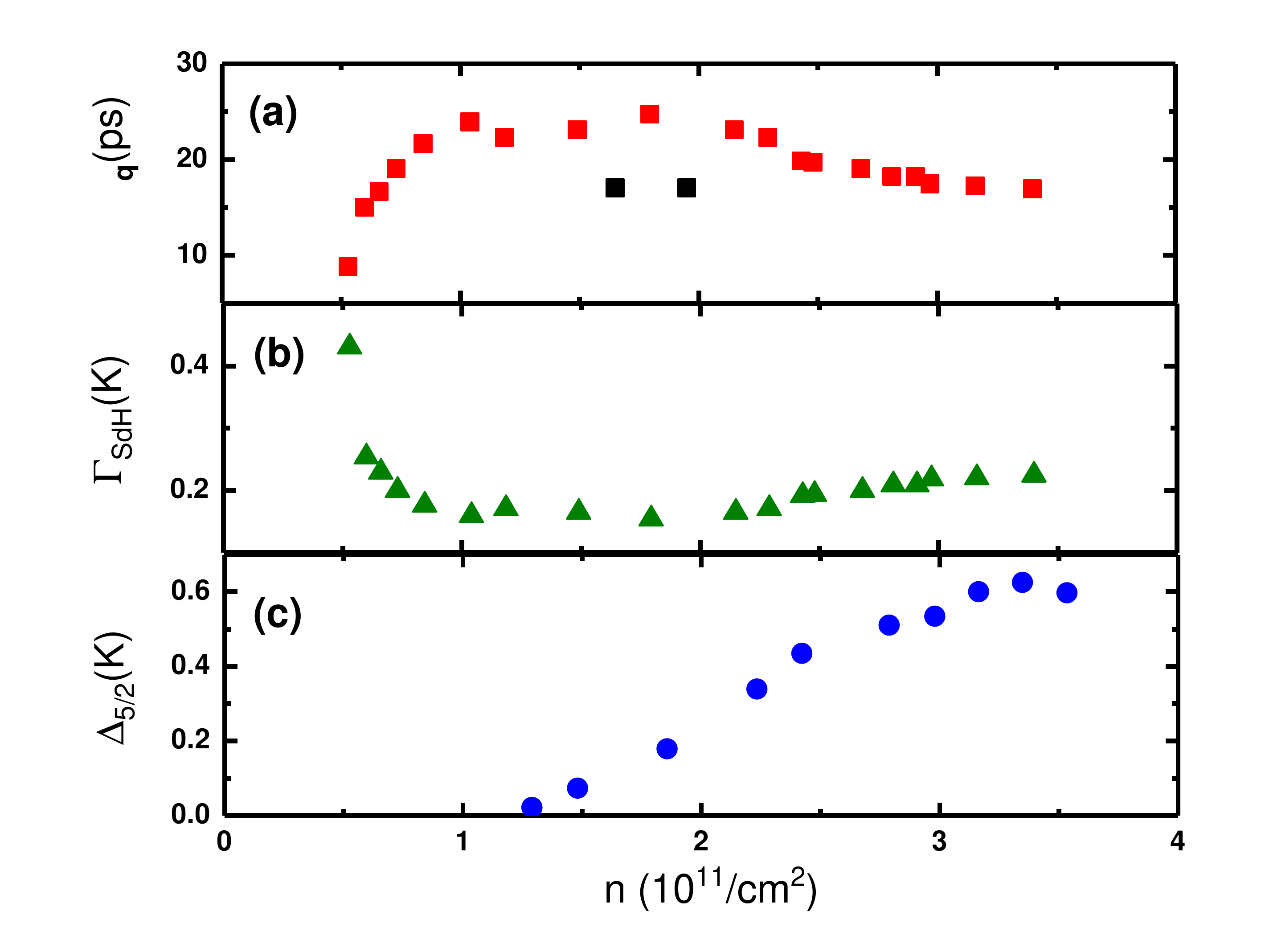}
\caption{\label{tauq} (color online). Characteristic properties of the back-gated sample as a function of
the electron density n. (a) Quantum lifetime $\tau_q$ extracted from Shubnikov de-Hass oscillations measured at T = 0.3K, both in dark (black) and after illumination (red).  (b) The Landau level broadening based on low-field quantum lifetime time with illumination, $\Gamma_{SdH}=\hslash/2\tau_{q}$. (c) Gap energy for the $\nu=5/2$ FQHS with illumination.}
\end{figure}

We extract $\tau_q$ from one of the back-gated samples at various densities after accounting for the effect of density inhomogeneity; the results are displayed in Fig. \ref{tauq}a. We observe that
the quantum lifetime initally increases monotonically from near depletion to n$\sim$1$\times$ $ 10^{11}/cm^{2}$, but it remains constant at around 25ps from n$\sim$1$\times$ $ 10^{11}/cm^{2}$ to n$\sim$2.5$\times$ $ 10^{11}/cm^{2}$ and starts to decrease slightly when density is higher then 2.5$\times$ $ 10^{11}/cm^{2}$. A similar trend was observed in Ref. \cite{Neubler2010}. The decrease in $\tau_q$ at high density is somewhat surprising since, assuming that the distribution of impurities remains the same, $\tau_q$ would be expected to increase monotonically with density in a gated device due to the increase of the Fermi wavevector $k_F$, as calculated in Ref. \cite{DasSarma2014}. This can be resolved by noting that, in our device, the gate consists of a heavily Si-doped GaAs layer, and when the density is increased by applying increased positive bias to the gate, an equal number of positively-charged impurities are ionized on the gate to maintain charge neutrality. These ionized donors act as scattering sites, and since their concentration increases linearly with density, $\tau_q$ starts to decrease at high densities. At low density, on the other hand, scattering is dominated by uniformly-distributed background impurities and fixed charged impurities in the doping well, and $\tau_q$ increases monotonically in that range; between the low and high density ranges, $\tau_q$ plateaus before turning over. These results are consistent with the model presented in Ref. \cite{DasSarma2014} in which the background impurity concentration is fixed but the remote ion impurity concentration is proportional to carrier density, as would be expected for the type of gate used in our device.

Most of the data we show is taken after the sample is illuminated with a red LED; this procedure is known to improve sample quality in terms of the strength of the FQHSs \cite{Gamez}. However, in Fig. \ref{tauq}a we also show two representative data points taken with no illumination of the sample. $\tau_q$ is lower without illumination, indicating that illumination improves screening of remote impurities that determine $\tau_q$. This effect is interesting as the change in $\tau_q$ does not accompany an increase in 2DEG density.  The illumination and subsequent relaxation simply allows the system to equilibrate to a configuration in which scattering is reduced.

Having acccounted for the impact of density inhomogeneity on $\tau_q$ we turn now to the relationship between $\tau_q$ and $\Delta_{5/2}^{meas}$, where $\Delta_{5/2}^{meas}$ is the experimentally measured gap, and discuss if $\tau_q$ can be used as a metric of quality relevant to N=1 LL. As seen in Fig. \ref{tauq}c, $\Delta_{5/2}^{meas}$ increases nearly monotonically with density. Clearly $\Delta_{5/2}^{meas}$ and $\tau_q$ behave differently as a function of density; a concomitant increase in $\tau_q$ is not observed in the density regime above n=1x10$^{11}$cm$^{-2}$. We note that our observation that $\Delta_{5/2}^{meas}$ increases with density while $\tau_q$ decreases is in direct contradiction with the expectations of Ref. \cite{DasSarma2014}. The simplest explanation for this is that the explicit increase of the intrinsic gap with density leads to the increase of the experimentally measured gap, and the effect of decreasing $\tau_q$ is overwhelmed. However, it is also possible that $\tau_q$ is simply not sensitive to the disorder relevant to $\Delta_{5/2}^{meas}$; in either case, $\tau_q$ cannot be used in a simple manner to predict $\Delta_{5/2}^{meas}$ without additional analysis in density-tunable devices.

We convert quantum lifetime to the Landau level broadening
using $\Gamma_{SdH}=\hslash /2\tau_{q}$ as shown in Fig. \ref{tauq}b. $\Gamma_{SdH}$ is usually interpreted as the magnetic field-independent energy broadening of the Landau levels. Then one would expect the relationship $\Delta_{5/2}^{theor}-\Delta_{5/2}^{meas}=\Gamma_{5/2}=\Gamma_{SdH}$ \cite{Chang1983}, where $\Delta_{5/2}^{theor}$ is the intristic gap in the absence of disorder. $\Delta_{5/2}^{theor}$ in this density range was numerically calculated in Ref. \cite{Neubler2010}, taking into account the finite width of the quantum well and LL mixing \cite{Morf2003}. According to Ref. \cite{Neubler2010} $\Delta_{5/2}^{theor}$ should exceed 2K at $n$ = 3.0x10$^{11}$cm$^{-2}$, far above the maximal value of $\Delta_{5/2}^{meas}$=0.625K. Clearly $\Gamma_{SdH}$ severely underestimates the level broadening $\Gamma_{5/2}$ relevant to the $\nu = 5/2$ state. This observation is consistent with other experiments reported previously \cite{Neubler2010, Nodar}. However we must be cognizant of the limitations of this analysis. Since the experimentally measured values of $\Delta_{5/2}^{meas}$ are much  smaller than the numerically calculated values it follows that small errors in the numerically calculated gap $\Delta_{5/2}^{theor}$ can lead to large changes of $\Gamma_{5/2}$ vs. density, including even its functional density dependence.  We also cannot completely rule out the possibility that $\Gamma_{5/2}$ is proportional to $\Gamma_{SdH}$ but differs by a scale factor. Regardless, we are led to the same conclusions as before: $\tau_q$ does not correlate directly with the gap $\Delta_{5/2}^{meas}$ nor can the disorder broadening of the $\nu = 5/2$ state be simply calculated from the expression $\Gamma=\hslash /2\tau_{q}$. In Ref. \cite{Qian}, we describe the utility of a different metric of 2DEG quality at T = 0.3K, $\rho_{5/2}$, the high temperature resistivity at $\nu=5/2$ where the state is best described as a Fermi sea of composite fermions. $\rho_{5/2}$ does show correlation with $\Delta_{5/2}$.

In conclusion, we consistently find that small density inhomogeneities in samples whose scattering from remote ionized impurities has been minimized yield Dingle plots that are non-linear and underestimate $\tau_q$.
This effect becomes significant with larger $\tau_{q}$ and lower temperature, as both move the onset of SdH oscillations to lower magnetic field where small density fluctuations have a larger impact. We have developed a method to incorporate this small density inhomogeneity by assuming a Gaussian distribution of 2DEG density and extract the intrinsic quantum lifetime using this method. We observe no correlation between $\tau_q$ and $\Delta_{5/2}$ in our density tunable devices, and conclude that $\tau_q$ is not useful for predicting the strength of the $\nu = 5/2$ FQHS.

This work was supported by the Department of Energy, Office of Basic Energy Sciences, under Award number DE-SC0006671.  Additional support for sample growth from the W. M. Keck Foundation and Microsoft Station Q is gratefully acknowledged.  SL and JF were supported by the Canada First Research Excellence Fund, QMI, and NSERC.


\begin{thebibliography}{100}
\bibitem{P} The densities are discretized and evenly spaced by $\Delta d=6\Delta n/$(2N+1), where N is the number of discrete density values included in the calculation (typically 100 in our simulations). The weight of density $n_{i}$ is given by $P(n_i)=\int_{n_i-\Delta d/2}^{n_i+\Delta d/2} g(n_{i}) dn_i$.
\end{thebibliography}
\end{document}